\documentclass[onecolumn, pra,showpacs,superscriptaddress]{revtex4}
\usepackage{graphicx}
\usepackage{dcolumn}
\usepackage{bm}
\usepackage{amsmath}
\begin{document}

\title{Dynamics of spinor Bose-Einstein condensate subject to dissipation
\thanks{Project supported by the National Natural Science Foundation of China (Grant No. 11004007)
and the Fundamental Research Funds for the Central Universities of China. Hao Y thanks the helpful discussion with Zhang Wen-Xian and Gu Qiang.}}


\author{Pang Man-Man and \ Hao Ya-Jiang$^{\rm }$\thanks{Corresponding author. E-mail:~haoyj@ustb.edu.cn}\\
$^{}${Department of Physics, University of Science and Technology Beijing, 100083 Beijing, P R China.
}}


\date{\today}

\begin{abstract}
We investigate the internal dynamics of the spinor Bose-Einstein Condensates subject to dissipation by solving the Lindblad master equation. It is shown that for the condensates without dissipation its dynamics always evolve along specific orbital in the phase space of ($n_0$, $\theta$) and display three kinds of dynamical properties including Josephson-like oscillation, self-trapping-like oscillation and 'running phase'. In contrast, the condensates subject to dissipation will not evolve along the specific dynamical orbital. If component-1 and component-(-1) dissipate in different rates, the magnetization $m$ will not conserve and the system transits between different dynamical regions. The dynamical properties can be exhibited in the phase space of ($n_0$, $\theta$, $m$).
\end{abstract}


\maketitle

\section{Introduction}
As Bose-Einstein condensates (BECs) are confined in an optical trap regardless of hyperfine state\cite{1,2}, the atomic spin degree of freedom are liberated and the spinor BECs are realized. This allow us to explore the properties related with spin of ultracold quantum gas\cite{3,4}. Magnetism of the condensates has ever been extensively investigated since its importance in traditional condensed matter physics \cite{5,6,7,8}. Ryan etc.have found that many body states of spinor atoms can be classified into several kinds of novel phases according to its spin symmetry\cite{9}. In addition, the realization of spinor BECs stimulated a great many of theories and experiments study on the dynamical properties of spin-dependent interaction. The researchers have investigated the irregular many-body spin-mixing dynamics of $F=$1 spinor condensates in the absence\cite{10,11,12,13} and in the presence of an external magnetic field\cite{14,15}. The magnetic properties of spinor BEC of high spin\cite{8} are also the present popular topics.

In experiment, one non-negligible problem is the coupling of condensates with the environment. For example, the interaction between the condensed atom and noncondensed thermal atoms results in the unavoidable atom loss\cite{16}. In some situation the dissipation effects\cite{17,18,19,20}, the thermal fluctuations\cite{21} and the dephasing\cite{22} will play crucial roles. S. Diehl has suggested that with the quantum optics method we can drive an open ultracold atomic system into a given pure quantum state, which provides a route towards preparing many-body states and non-equilibrium quantum phases\cite{23}. The proposal to preparing spin squeezed state, phase- and number-squeezed state has been given\cite{24,25}.

In this paper, we investigate the spin-mixing dynamics of spin-1 antiferromagnetic spinor BECs subject to dissipation. As the dissipation rate are different for each component, the magnetization will not conserved and the pseudo-angular momentum operator defined in Ref. \cite{26,27} cannot describe the dynamics of spinor BECs. By constructing a set of operators, the Lindblad master equation\cite{28,29}, which govern the dynamics of the open quantum system, can be formulated as a group of nonlinear dynamical evolution equations under the mean-field approximation. Therefore, we can obtain the dynamics of open spinor BECs by numerically solving the set of nonlinear equations.

\section{Model and method}
We consider spin-1 spinor BECs with $N$ atoms of mass $M$ trapped in a spin-independent external potential $V(\bm r)$, the second quantized Hamiltonian is formulated as
\begin{equation*}
 \mathcal {H}= \int d\bm r\hat{\Psi}_{\alpha}^{\dag}(\bm r)\left[-\frac{\hbar^2\nabla^{2}}{2M}+V(\bm r)-\mu\right]\hat{\Psi}_\alpha(\bm r)+\frac{c_0}{2}\hat{\Psi}_{\alpha}^{\dag}(\bm r) \hat{\Psi}_{\beta}^{\dag}(\bm r)\hat{\Psi}_{\beta}(\bm r)\hat{\Psi}_{\alpha}(\bm r) +\frac{c_2}{2}\hat{\Psi}_{\alpha}^{\dag}(\bm r) \hat{\Psi}_{\alpha\prime}^{\dag}(\bm r){\bf F}_{\alpha \beta} \cdot {\bf F}_{\alpha \prime \beta \prime}\hat{\Psi}_{\beta \prime}(\bm r)\hat{\Psi}_{\beta}(\bm r), \tag{$1$}
 \end{equation*}
where $\hat{\Psi}_{\alpha}(\bm r)\left(\alpha=0,\pm1\right)$ is the field operator associated with the annihilation of atom in the hyperfine spin state $\left|F=1,{{m}_{F}}=\alpha \right\rangle$. The interaction parameters $c_0=4\pi \hbar ^2(a_0+2a_2)/(3M)$ and $c_2=4\pi\hbar^2(a_2-a_0)/(3M)$ correspond to the spin-independent and spin-dependent interaction, respectively. Here $a_f$ ($f=0,2$) is the $s$-wave scattering length for atoms in the channel of total spin $f$. Because of $|c_2|\ll c_0$ ($a_0=50a_B$ and $a_2=55a_B$ for $^{23}$Na with $a_B$ being Bohr radius, the single-mode approximation (SMA)\cite{30} was utilized effectively to study spinor BECs. In the approximation the spatial wave functions for each spin component $\phi_{\alpha}(\bm r)$ ($\alpha=0,\pm 1$) are described by the same wave function $\phi(\bm r)$ determined by the ground-state solution of the Gross-Pitaevskii equation\cite{31,32} rather than the coupled Gross-Pitaevskii equations:
 \begin{equation*}
 \ (-\frac{\hbar^2\nabla^{2}}{2M}+V(\bm r)+c_0N|\phi|^{2})\phi=\mu\phi, \tag{$2$}
 \end{equation*}
with $\mu$ being chemical potential. Here $N$ is the total particle number, which is a classical number. Therefore we have $\hat{\Psi}_{\alpha}\approx\hat{a}_{\alpha}\phi(\bm r)$ with $\hat{a}_{\alpha}$ being annihilation operator for $\alpha$ component, and the Hamiltonian is given by
\begin{align}
\hat{H}&=\mu\hat{N}-\lambda^{\prime}_{0}\hat{N}(\hat{N}-1)\nonumber\\
&{} +\lambda_{2}^{\prime}(\hat{a}_{1}^{\dag}\hat{a}_{1}^{\dag}\hat{a}_{1}\hat{a}_{1}
 +\hat{a}_{-1}^{\dag}\hat{a}_{-1}^{\dag}\hat{a}_{-1}\hat{a}_{-1}
 +2\hat{a}_{1}^{\dag}\hat{a}_{0}^{\dag}\hat{a}_{1}\hat{a}_{0} {}\nonumber\\
 &{}+2\hat{a}_{-1}^{\dag}\hat{a}_{0}^{\dag}\hat{a}_{-1}\hat{a}_{0}
 -2\hat{a}_{1}^{\dag}\hat{a}_{-1}^{\dag}\hat{a}_{1}\hat{a}_{-1}
 +2\hat{a}_{0}^{\dag}\hat{a}_{0}^{\dag}\hat{a}_{1}\hat{a}_{-1}+2\hat{a}_{1}^{\dag}\hat{a}_{-1}^{\dag}\hat{a}_{0}\hat{a}_{0}), \tag{$3$}
\end{align}
where $\hat{N}$ is the total atom number operator in the condensate and $2\lambda^{\prime} _{i}\equiv c_{i}\int (|\phi(\bm r)|^{4})d\bm r$ ($i=0,2$).

For a closed system, the total atom number $\hat N$ and magnetization $\hat m=\hat a_1^\dagger\hat a_1 -\hat a_{-1}^\dagger\hat a_{-1}$ are conserved quantities. Following the algebra in Ref. \cite{26,27}, we can define the operators $\hat L_-=\sqrt 2(\hat a_1^\dagger\hat a_0 +\hat a_0^\dagger\hat a_{-1})$, $\hat L_+=\sqrt 2(\hat a_0^\dagger\hat a_1 +\hat a_{-1}^\dagger\hat a_0)$, $\hat L_z=\sqrt 2(\hat a_{-1}^\dagger\hat a_{-1} -\hat a_1^\dagger\hat a_1)$, which satisfy the usual angular momentum commutation relation. They can completely describe the properties of the spin-1 BECs if the system is closed. In the present paper we will investigate the dynamics of spinor BECs subject to dissipation for the coupling with the environment. In this case both the total atom number and magnetization will not be conserved and the new algebraic structure have to be employed. We construct the operators $\hat Y_k=i(\hat{a}_{\alpha}^{\dag}\hat{a}_{\beta}-\hat{a}_{\beta}^{\dag}\hat{a}_{\alpha})$ and $\hat Z_k=(\hat{a}_{\alpha}^{\dag}\hat{a}_{\beta}+\hat{a}_{\beta}^{\dag}\hat{a}_{\alpha})$ ($(\alpha,\beta)$ is (1,-1), (-1,0) and (0,1) for $k=1,2,3$, respectively). Combing $\hat Y_k$ and $\hat Z_k$ with $\hat N$, $\hat m$, and atom number in 0-component $\hat N_0$, the open system can be described.
With the help of the above operators, $\hat H$ takes the following form
\begin{equation}
  \hat{H}=\mu\hat{N}-\lambda_{0}^{\prime}\hat{N}(\hat{N}-1)+\frac{1}{2}\lambda_{2}^{\prime}[(\hat{Z}_{2}+\hat{Z}_{3})^{2}+(\hat{Y}_{2}+\hat{Y}_{3})^{2}
      +2\hat{m}^{2}-4\hat{N}].\tag{$4$}
 \end{equation}

For the system coupled with environment inducing a non-equilibrium dynamics, its time evolution is determined by the Lindblad master equation\cite{29} for the reduced system density operator $\hat {\rho}$
  \begin{align}
   \dot{\hat{\rho}}=& \frac1{i\hbar}[\hat{H},\hat{\rho}]-\frac{1}{2}\sum\limits_{\alpha=0,\pm1}\gamma_{a_{\alpha}}(\hat{a}_{\alpha}^{\dag}\hat{a}_{\alpha}\hat{\rho}
             +\hat{\rho}\hat{a}_{\alpha}^{\dag}\hat{a}_{\alpha}-2\hat{a}_{\alpha}\hat{\rho}\hat{a}_{\alpha}^{\dag}).  \tag{$5$}
  \end{align}
Here, $\hat H$ is the system Hamiltonian, and $\gamma_{a_{\alpha}}$ is the dissipation rate of component $\alpha$ determined by the interaction between the cold atoms and thermal atoms. The one-time average of arbitrary operator $\hat A$ can be calculated via $<A>=\text{ tr}[\hat{A}\hat{\rho }(t)]$ and its time derivative $<\dot{\hat A}>=\text{tr}[\hat{A}\dot{\hat{\rho }}(t)]$. We define $n(t)=\text {tr}[\hat N \hat {\rho}(t)]/N$, $m(t)=\text{ tr}[\hat m \hat {\rho}(t)]/N$, $n_0(t)=\text{ tr}[\hat N_0 \hat {\rho}(t)]/N$, $y_i(t)=\text{ tr}[\hat Y_i \hat {\rho}(t)]/N$, $z_i(t)=\text{ tr}[\hat Z_i \hat {\rho}(t)]/N$. Inserting the Hamiltonian $\hat H$ into the Eq.(5) and taking the first-order approximation, i.e. $<\hat A \hat B>=<\hat A><\hat B>$, we can obtain a set of closed equations
\begin{align}
\dot{n}=&-\frac{1}{2}(\gamma_{a_{1}}+\gamma_{a_{-1}})n-\frac{1}{2}(\gamma_{a_{1}}-\gamma_{a_{-1}})m+\frac{1}{2}(\gamma_{a_{1}}+\gamma_{a_{-1}}-2\gamma_{a_{0}})n_{0},\nonumber\\
\dot{m}=&-\frac{1}{2}(\gamma_{a_{1}}-\gamma_{a_{-1}})n-\frac{1}{2}(\gamma_{a_{1}}+\gamma_{a_{-1}})m+\frac{1}{2}(\gamma_{a_{1}}-\gamma_{a_{-1}})n_{0},\nonumber\\
\dot{n}_{0}=&2\lambda^{\prime}_{2}(y_{2}z_{3}-y_{3}z_{2})-\gamma_{a_{0}}n_{0},\nonumber\\
 \dot{y}_{1}=&\lambda^{\prime}_{2}\left(y_{2}^{2}-y_{3}^{2}-z_{2}^{2}+z_{3}^{2}-4mz_{1}\right)-\frac{1}{2}\left(\gamma_{a_{1}}+\gamma_{a_{-1}}\right)y_{1},\nonumber\\
\dot{y}_{2}=&\lambda^{\prime}_{2}[-y_{1}y_{3}-y_{1}y_{2}+z_{1}z_{2}+z_{1}z_{3}+z_{2}(n-m-3n_{0})+z_{3}(n   {}\nonumber\\
          &{}-m-3n_{0})+2mz_{2}]-\frac{1}{2}(\gamma_{a_{-1}}+\gamma_{a_{0}})y_{2},\nonumber\\
\dot{y}_{3}=&\lambda^{\prime}_{2}[y_{1}y_{3}+y_{1}y_{2}-z_{1}z_{2}-z_{1}z_{3}-z_{2}(n+m-3n_{0})-z_{3}(n {}\nonumber\\
           &{} +m-3n_{0})+2mz_{3}]-\frac{1}{2}(\gamma_{a_{0}}+\gamma_{a_{1}})y_{3},\nonumber    \\
\dot{z}_{1}=&\lambda^{\prime}_{2}(-2y_{2}z_{2}+2y_{3}z_{3}+4my_{1})-\frac{1}{2}(\gamma_{a_{1}}+\gamma_{a_{-1}})z_{1},\nonumber\\
\dot{z}_{2}=&\lambda^{\prime}_{2}[y_{1}z_{2}+y_{1}z_{3}+y_{2}z_{1}-y_{2}(n-m-3n_{0})+y_{3}z_{1}-y_{3}(n {}\nonumber\\
             &{}-m-3n_{0})-2my_{2}]-\frac{1}{2}(\gamma_{a_{0}}+\gamma_{a_{-1}})z_{2},\nonumber
\end{align}
\begin{align}
\dot{z}_{3}=&\lambda^{\prime}_{2}[-y_{1}z_{2}-y_{1}z_{3}-y_{2}z_{1}+y_{2}(n+m-3n_{0})-y_{3}z_{1}+y_{3}(n {}\nonumber\\
             &{}+m-3n_{0})-2my_{3}]-\frac{1}{2}(\gamma_{a_{0}}+\gamma_{a_{1}})z_{3}.\tag{$6$}
\end{align}
For simplicity, we use the dimensionless formula and in the following evaluation the time and energy are in units of the trapping frequency $\omega^{-1}$ and $\hbar \omega$, respectively.

After solving the above equations, we can calculate the interested quantities. To compare with the dynamical properties of closed spinor BECs, we will focus on the evolving dynamics in phase space of $(n_0,\theta)$. $\theta$ is the relative phase of three components $\theta_\alpha$ ($\theta=\theta_{1}+\theta_{-1}-2\theta_{0}$). The evolution of $n_0$ can be obtained by solving the equations. As we have time dependence of $\theta$, the dynamical orbitals are determined in the phase space of $(n_0,\theta)$. According to Ref. \cite{33}, in the single-mode approximation we have $\theta=\arctan\frac{y_{3}z_{2}-y_{2}z_{3}}{y_{2}y_{3}+z_{2}z_{3}}$. While the population in component 1 and -1 can be obtained by ${n}_{1}=\frac{{n}+{m}-{n}_{0}}{2}$ and ${n}_{-1}=\frac{{n}-{m}-{n}_{0}}{2}$.

\section{Dynamics of Open Spinor BECs}
In this section we first show the dynamical properties of spinor BECs decoupled with environment, i.e., there is no particle dissipation, and then we exhibit the dynamics as each component is subject to dissipation. In the present paper we will display our results by taking the antiferromagnetic spinor BECs as the example and the spin-dependent interaction constant $\lambda_{2}^{\prime}=0.2$.

\begin{center}
\includegraphics[width=7cm]{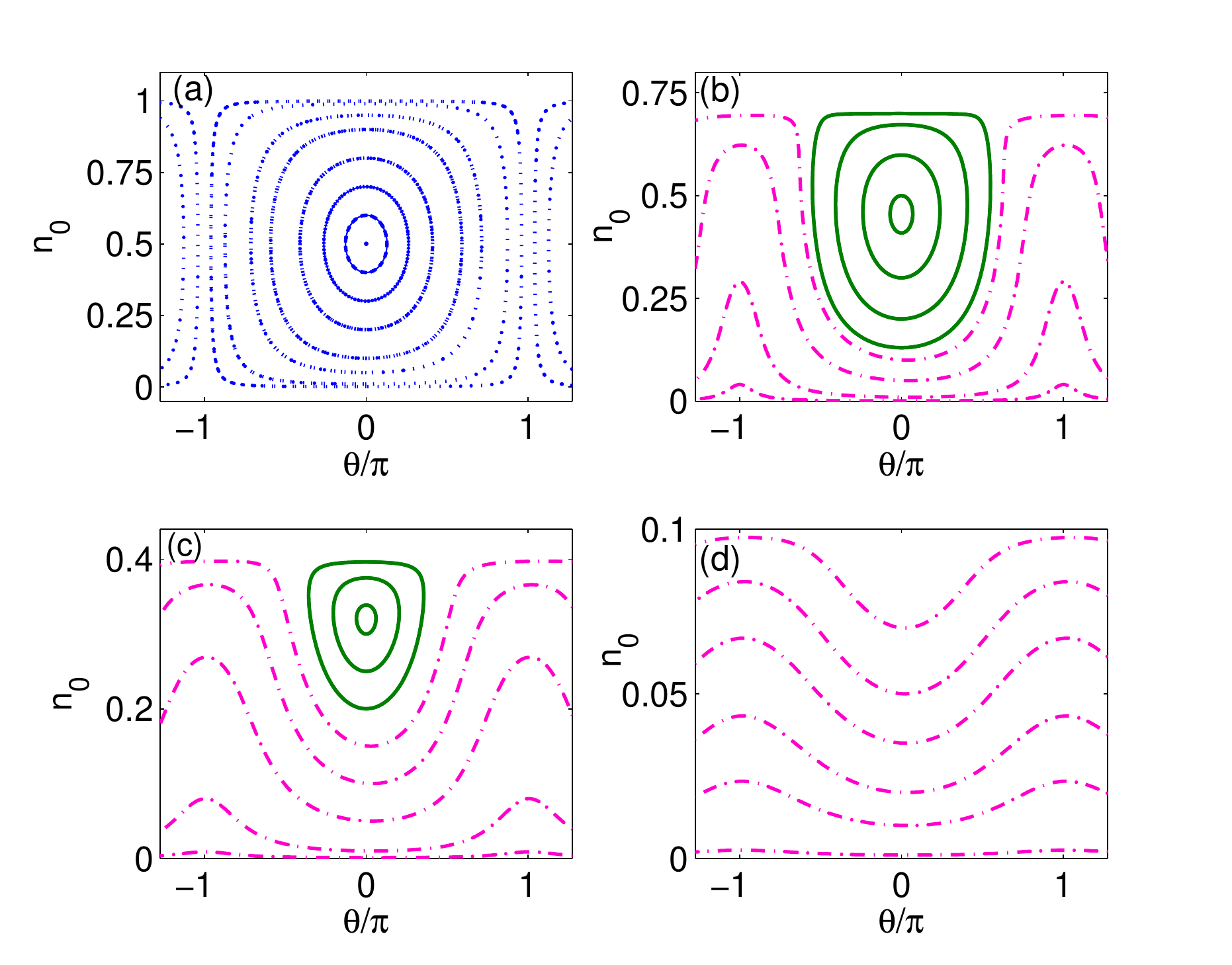}\\[5pt]  
\parbox[c]{15.0cm}{\footnotesize{\bf Fig.~1.} (color online) Dynamical trajectories in $n_0-\theta$ plane for $\gamma_{a_{1}}=\gamma_{a_{-1}}=\gamma_{a_{0}}=0$: (a) $m$=0; (b) $m$=0.3; (c) $m$=0.6; (d) $m$=0.9.
}
\end{center}
In Fig. 1 the dynamical orbitals of close spinor BECs are shown for different magnetization $m$. For given initial magnetization $m(0)$, it is conserved and will not change in the full dynamics. This is consistent with the fact that magnetization is a conserved quantity in the spinor BECs. For different magnetization the dynamical orbitals display specific characteristic even for the same initial condition. In Fig. 1a ($m=0$), the dynamical trajectories evolve in the center of $n_0=0.5$ and $\theta=0.0$ and all of them are closed. As a corresponding to the two component BECs or BECs in double well\cite{34}, we define particle number difference between the compoent-0 and the others, $z=(n_1+n_{-1})-n_0$. The mean value of $z$ in an evolving period is 0.0 ($\bar n_0=0.5$), which is similar to the Josephson oscillation in two component BECs. In Fig.1b and Fig.1c ($m=$0.3 and 0.6) the open orbitals appear. In these situations, with the time evolution the phase $\theta$ will be always 'running' rather than be periodical. The orbitals have the characteristic of 'running phase'. For the closed dynamical orbitals here, $\bar n_0$ deviate from 0.5 ($\bar z \ne 0.0$), which are much like to the 'self-trapping' in two component BECs. Similar to the case of two component BECs, three kinds of dynamical orbitals are shown for different initial population ($n_0(0)$ and $m(0)$) in spinor BECs, Josephson-like oscillation, self-trapping-like, and 'running phase'. For large magnetization the dynamical orbitals show the 'running phase' properties ($m$=0.9 in Fig. 1d).

For the given magnetization $m$ the spinor BECs is always in the specified dynamical region in the phase space of ($n_0, \theta$). The system will evolve in one special dynamical orbital and keep up evolving with $m$ being constant. This is what we have shown so far in Fig.1, all of which are internal spin-mixing dynamics of spinor BECs. As the spinor BECs subject to dissipation, the internal dynamical properties will be affected by the external environment. The magnetization $m$ will change if the dissipation rates of component-1 and component-(-1) are distinct. Even if the dissipation rates are same for components 1 and -1, the dynamical properties will extremely different from the spinor BECs without coupling with the environment.

\begin{center}
\includegraphics[width=7cm]{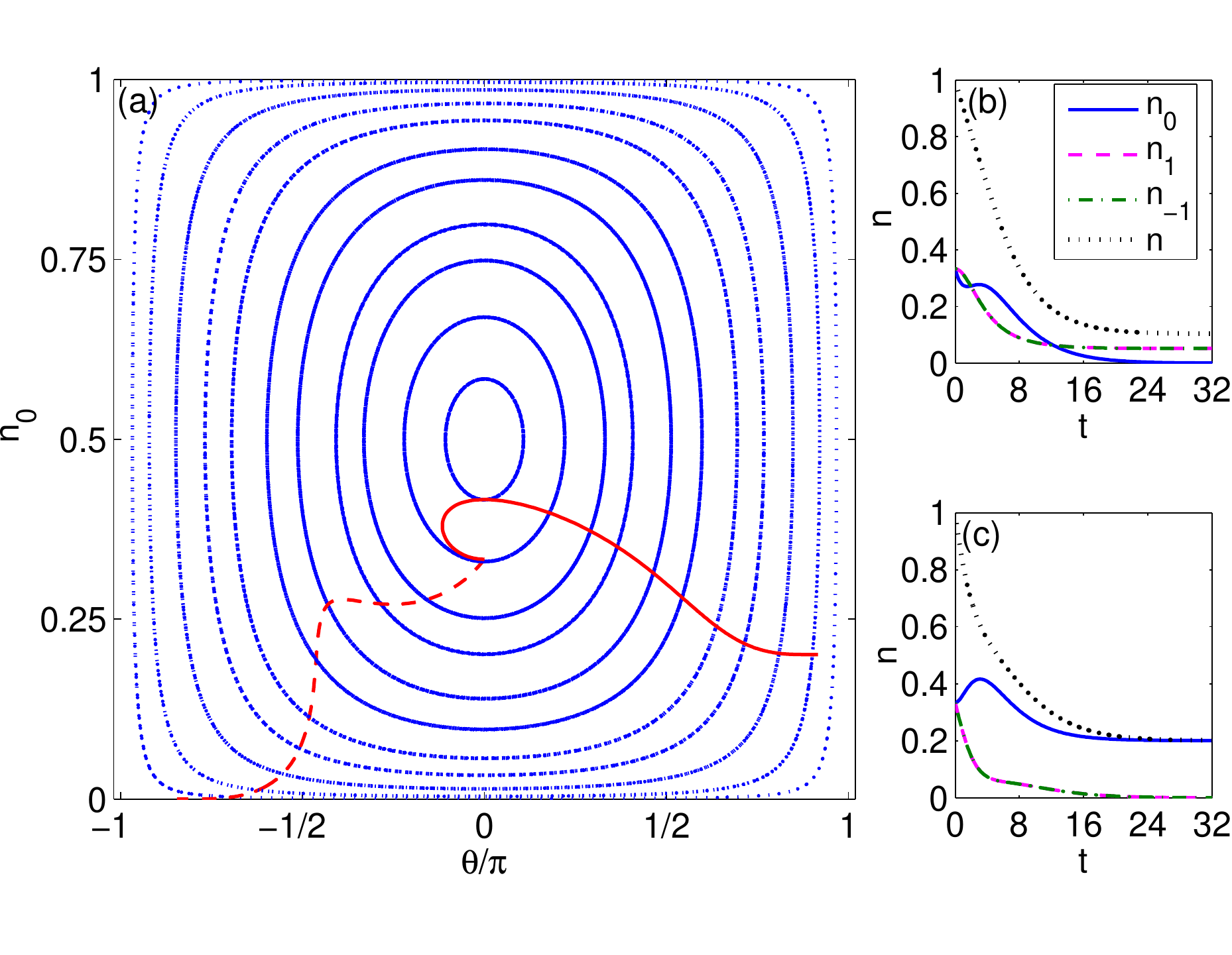}\\[5pt]  
\parbox[c]{15.0cm}{\footnotesize{\bf Fig.~2.} (color online) (a) Dynamical orbitals of open spinor BECs of $m$=0. Dashed lines: $\gamma_{a_{1}}=\gamma_{a_{-1}}=0,\gamma_{a_{0}}={1}/{3}$; Solid line: $\gamma_{a_{1}}=\gamma_{a_{-1}}={1}/{3},\gamma_{a_{0}}=0$. To compare with the close system, the dynamical orbitals of the system without dissipation are shown (dotted lines). (b) The dynamical evolution of the particle number of each components for $\gamma_{a_{1}}=\gamma_{a_{-1}}=0,\gamma_{a_{0}}=1/3$. (c) The dynamical evolution of the particle number of each components for $\gamma_{a_{1}}=\gamma_{a_{-1}}=1/3,\gamma_{a_{0}}=0$.}
\end{center}
In Fig. 2 the dynamical orbitals are shown for the case of $\gamma_{a_1}=\gamma_{a_{-1}}$ and $m=0$. Since the atoms in component-1 and component-(-1) dissipate in the same rate and the internal dynamical properties of spinor BECs, the magnetization will conserved as a constant 0. We can exhibit the dynamical orbitals of closed spinor BECs in the phase space of ($n_0$, $\theta$) of $m=0$. It is shown that the system still evolves in Josephson-like oscillation region, but it will transit between different orbitals rather than evolve in one specific close orbital. In addition, the external coupling with environment can be utilized to control the evolving orbital and the arrived final state. As only component-0 dissipate ($\gamma_{a_1}=\gamma_{a_{-1}}=0$ and $\gamma_{a_0}=1/3$), the atoms of component-0 will completely disappear and some atoms of component-1 and component-(-1) remain in the system ($n_1=n_{-1}=0.05$ finally). As only component-1 and component-(-1) dissipate ($\gamma_{a_1}=\gamma_{a_{-1}}=1/3$ and $\gamma_{a_0}=0$), the contrary situation will take place ($n_0$=0.2). In this case most atoms of component-0 remain in the system after the other two component loss completely.


\begin{center}
\includegraphics[width=7cm]{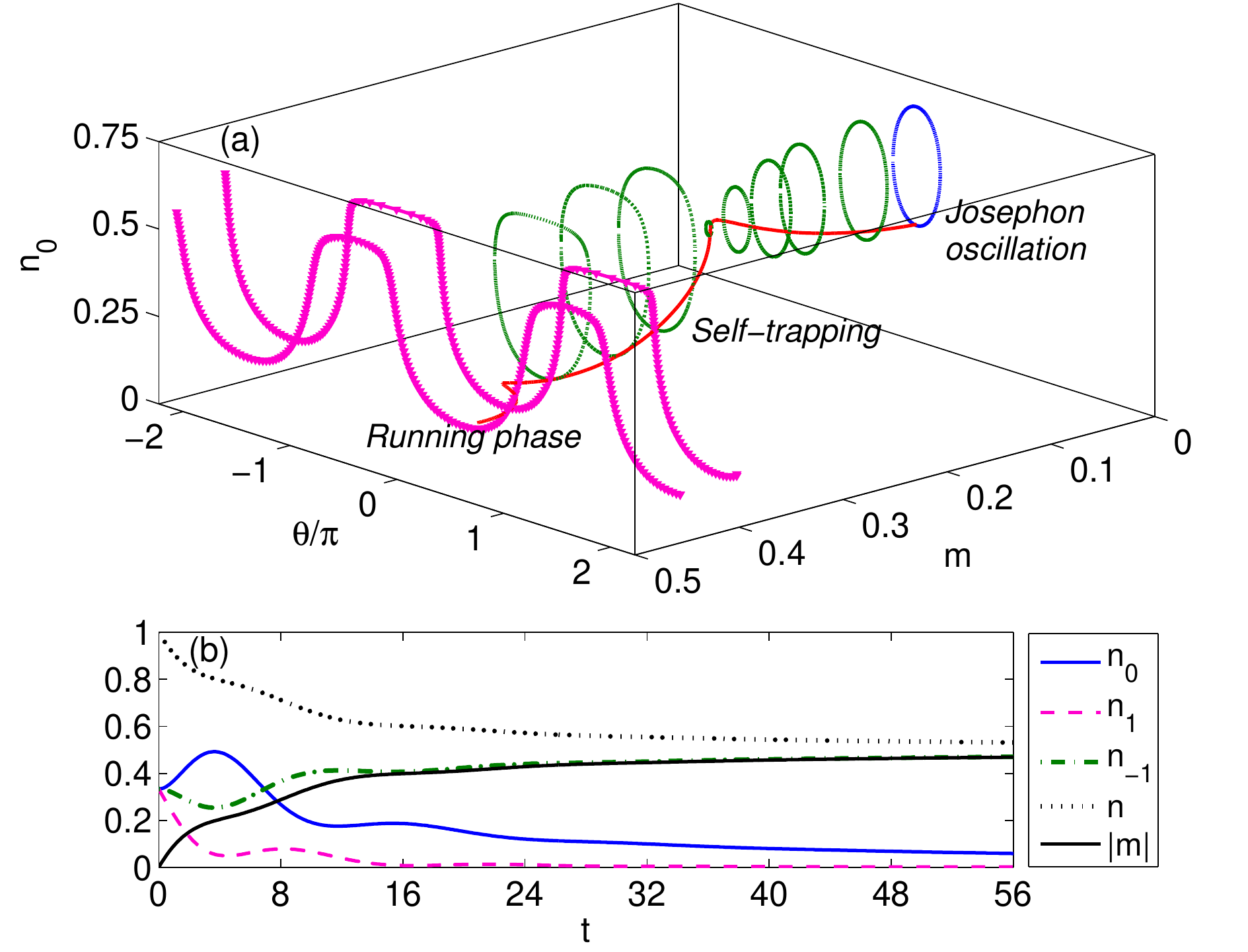}\\[5pt]  
\parbox[c]{15.0cm}{\footnotesize{\bf Fig.~3.} (color online) (a) Dynamical orbitals of open spinor BECs. To compare with the close system, the dynamical orbitals of the system without dissipation are shown (dotted lines). $\gamma_{a_{1}}={1}/{3}$,$\gamma_{a_{-1}}=\gamma_{a_{0}}=0$.
(b) The dynamical evolution of the particle number of each components.}
\end{center}

In Fig. 3a the dynamical orbitals are shown for the case of $\gamma_{a_1}\neq\gamma_{a_{-1}}$ and $m(0)=0$. As the dissipation rate of component-1 and component-(-1) are different, the dynamical orbital will evolve in the phase space of ($n_0$, $\theta$, $m$) rather than ($n_0$, $\theta$) because the magnetization will change in this case. The dynamics evolve starts at the Josephson-like oscillation regime. If the spinor BECs do not subject to dissipation, the system maintains the stable Josephson-like oscillation along the specific close orbital of $m=0$. Since the component-1 dissipate in the rate $\gamma_{a_1}=1/3$, the magnetization $m$ deviates from zero and the system gradually evolve into self-trapping-like region at $|m|$=0.1. With the loss of spin-1 atoms, the magnetization become larger further and the system evolve into the 'running phase' region at $|m|=0.4$. Therefore by controlling the component-dependent dissipation rate the system can evolve from the stable Josephson-like oscillation into self-trapping-like region and ultimately into the 'running phase' region.

\section{Conclusion}
In summary, we investigated the spin-mixing dynamics of a spinor BECs subject the dissipations by constructing a set of operators and solving the Lindblad master equation. As the system is closed, for different initial conditions the dynamical orbitals show the properties of Josephson-like oscillation, self-trapping, and 'running phase'. The spinor BECs evolve along the specific dynamical orbital. While the system is coupled with the environment and atom dissipation exist, the system will transit between different dynamical orbitals. For the spinor BECs with the same dissipation rates for component-$\pm 1$, the magnetization conserve and the system evolves in the phase space of ($n_0$, $\theta$). For the spinor BECs with different dissipation rates for component-$\pm 1$, the system can evolve from Josephson-like oscillation into self-trapping-like region and 'running phase' region. In this situation, the dynamical orbitals evolve in the phase space of ($n_0$, $\theta$, $m$).


\addcontentsline{toc}{chapter}{References}
\begin{thebibliography}{99}\footnotesize

\bibitem{1} Stamper-Kurn D M, Andrews M R, Chikkatur A P,Inouye S,Miesner H J,Stenger J and Ketterle W {1998 \text{Phys. Rev. Lett } \textbf{80} 2027}
\bibitem{2} Stenger J,Inouye S,Stamper-Kurn D M, Miesner H J,Chikkatur A P and Ketterle W  {1998 \emph{Nature} \textbf{396} 345}
\bibitem{3} Lamacraft A  {2007 \emph{Phys. Rev. Lett} \textbf{98} 160404}
\bibitem{4} Yuki K and Masahito U  {2012\emph{ Physics Reports} \textbf{520}  253¨C381}
\bibitem{5} Pu H, Zhang W P and Pierre M {2001\emph{Phys. Rev. Lett} \textbf{87} 140405}
\bibitem{6} Gu Q and Richard A K  {2003\emph{Phys. Rev. A} \textbf{68}  031604(R)}
\bibitem{7} Recati A,Fedichev P O,Zwerger W and Zoller P  {2003\emph{Phys. Rev. Lett} \textbf{90}  020401}
\bibitem{8} Pasquiou P,Mar\'{e}chal E,Bismut G,Pedri P,Vernac L,Gorceix O and Laburthe-Tolra B {2011\emph{Phys. Rev. Lett} \textbf{106}  255303}
\bibitem{9} Ryan B, Ari T and Eugene D {2006\emph{Phys. Rev. Lett} \textbf{97}  180412}

\bibitem{10} Chong G S and Borondo F {2008 \emph{Phys. Rev. E} \textbf{78} 016204}
\bibitem{11} Gu Q and Qiu H B {2007 \emph{Phys. Rev. Lett} \textbf{98} 200401}
\bibitem{12} Zhang J, Yang B G and Zhang Y B {2011 \emph{Phys. Rev. A} \textbf{83} 053634}
\bibitem{13} Cheng R, Liang J Q and Zhang  Y B {2005 \emph{J. Phys. B: At. Mol. Opt.Phys.} \textbf{38} 2569}


\bibitem{14} Romano D R and de Passos E J V {2004 \emph{Phys. Rev. A}\textbf{70} 043614}
\bibitem{15} Li H B, Pu Z G, Chapman  M S and Zhang W X {2015 \emph{Phys. Rev. A}\textbf{92} 013630}
\bibitem{16} Hao Y J and Gu Q {2011  \emph{Phys. Rev. A} \emph{83} 043620}
\bibitem{17} Witthaut D, Trimborn F and Wimberger S {2009 \emph{Phys. Rev. A} \textbf{79} 033621}
\bibitem{18} Trimborn F, Witthaut D and Wimberger S {2008  \emph{J. Phys. B: At. Mol. Opt. Phys.} \textbf{41} 171001}
\bibitem{19} Syassen N, Bauer D M, Lettner M, Volz T and Dietze D {2008 \emph{Science}  \textbf{320}  1329}
\bibitem{20} Valeriy A B, Vladimir V K, V\'{\i}ctor M.P\'{e}rez-Garc\'{\i}a  and Herwig Ott {2009 \emph{Phys. Rev. Lett } \textbf{102} 144101}
\bibitem{21} Rudolf G, Hemmerling B, Jonas F, Michael A and Markus K O {2006 \emph{Phys. Rev. Lett} {\bf  96} 130404}
\bibitem{22} Gati R, Estve J, Hemmerling B, Ottenstein T B, Appmeier J, Weller A and Oberthaler M K {2006  \emph{New J. Phys } \textbf{8} 189}
\bibitem{23} Diehl S ,Micheli A, Kantian A, Kraus B,B\"{u}chler H P and Zoller P {2008 \emph{Nature Physics} \textbf{4}   878-883}
\bibitem{24} Antonio D L {2013 \emph{Phys. Rev. Lett} \textbf{110}  120403}
\bibitem{25} Roland Cristopher F C, Sebastian D, Harri M, Markus O and Gentaro W {2014 \emph{Phys. Rev. A} \textbf{89}  013620}
\bibitem{26} Wu Y {1996 \emph{Phys. Rev. A} \textbf{54} 4534}
\bibitem{27} Law C K,Pu H and Bigelow N P {1998 \emph{Phys. Rev. Lett} \textbf{81} 5257}
\bibitem{28} Lindblad G {1976 \emph{Commun. Math. Phys.}  \textbf{48}  119}
\bibitem{29} Gardiner C W and Zoller P {2004 \emph{Quantum Noise},3rd edn. (Berlin Heidelberg:Springer-Verlag Press) pp.~13--18}
\bibitem{30} Yi S,M\"{u}stecapl{\i}o\v{g}lu\"{O}E, Sun C P and You L {2002 \emph{Phys. Rev. A} \textbf{66} 011601(R)}
\bibitem{31} Dalfovo F, Giorgini S, Pitaevskii L P and Stringari S {1999 \emph{Rev.Mod.Phys } \textbf{71} 463}
\bibitem{32} Pethick C J and Smith H {2008  \emph{Bose-Einstein Condensations in Dilute Gases} 2nd edn. (Cambridge:Cambridge University Press) pp.~159-162}
\bibitem{33} Black A T, Gomez E, Turner L D, Jung S and Lett P D {2007 \emph{Phys. Rev. Lett} \textbf{99} 070403}
\bibitem{34} Smerzi A, Fantoni S, Giovanazzi S and Shenoy S R {1998 \emph{Phys. Rev. Lett} \textbf{79} 4950}
\end{thebibliography}
\end{document}